\documentclass{article}

\usepackage{PRIMEarxiv}
\usepackage{amsmath}
\usepackage[round]{natbib}

\usepackage[utf8]{inputenc} % allow utf-8 input
\usepackage[T1]{fontenc}    % use 8-bit T1 fonts
\usepackage{hyperref}       % hyperlinks
\usepackage{url}            % simple URL typesetting
\usepackage{booktabs}       % professional-quality tables
\usepackage{amsfonts}       % blackboard math symbols
\usepackage{nicefrac}       % compact symbols for 1/2, etc.
\usepackage{microtype}      % microtypography
\usepackage{lipsum}
\usepackage{fancyhdr}       % header
\usepackage{graphicx}       % graphics
\graphicspath{{media/}}     % organize your images and other figures under media/ folder
\usepackage{verbatim}
\begin{document}

\title{Consistent time travel for realistic interactions with historical data: reinforcement learning for market making}
\author{Vincent Ragel and Damien Challet\thanks{corresponding author: \protect\href{mailto:}{damien.challet@centralesupelec.fr}.}\\
Université Paris-Saclay\\CentraleSupélec\\
Laboratoire de Mathématiques et Informatique pour la Complexité et les Systèmes\\
91192 Gif-sur-Yvette, France}
\maketitle
\begin{abstract}
Reinforcement learning works best when the impact of the agent's actions on its environment can be perfectly simulated or fully appraised from available data.
Some systems are however both hard to simulate and very sensitive
to small perturbations. An additional difficulty arises when a RL agent is trained offline to be part of a multi-agent system using only anonymous data, which makes it impossible to infer the state of each agent, thus to use data directly. Typical examples are competitive systems without agent-resolved data such as financial markets. We introduce consistent data time travel for offline RL as a remedy for these problems: instead of using historical data in a sequential way, we argue that one needs to perform time travel in historical data, i.e., to adjust the time index so that both the past state and the influence of the RL agent's action on the system coincide with real data. This both alleviates
the need to resort to imperfect models and consistently accounts for
both the immediate and long-term reactions of the system when using anonymous historical data. We apply this idea to market making
in limit order books, a notoriously difficult task for RL; it turns out that the gain of the agent is significantly higher with data time travel than with naive sequential data, which suggests that the difficulty of this task for RL may have been overestimated.
\end{abstract}

\section{Introduction}

Training an agent with Reinforcement Learning (RL) requires a faithful description of the interaction of the agent with its environment. This makes it possible to account  for the long and short term impact of the agent's actions, and thus to compute realistic rewards. 
Some systems are more amenable to RL than others: video games, and
by extension artificial worlds, offer a perfect environment simulator \citep{mnih2015human,schrittwieser2020mastering}. Most world simulators account for the agent's impact imperfectly, which may not matter much if the agent's impact is likely
to be small and short-lived. Historical data offers another way to perform offline training  (see \cite{levine2020offline} for a review).

RL is least effective when used in systems that are very sensitive to every single event with only partial information, because they are hard to simulate and publicly available data is incomplete. In other words, RL training may be so
imperfect as to lead the agent to learn erroneous wisdom from wrong rewards or obtaining systematically biased rewards, leading to overly optimistic or pessimistic outcomes.

Here, we tackle the offline training of a RL agent with only anonymous data for competitive, asynchronous, hyper-reactive multi-agent systems with long memory, action latency, and for which no good model exists. Specifically, we are interesting in high-frequency trading where the agents send orders or update them very frequently, but our approach generalises to other such multi-agent systems. 

High-frequency traders exchange assets via so-called order-driver markets, where a matching engine aggregates a sequence of orders to buy or to sell a given asset
at prices chosen by traders. Buy and sell orders that can be matched result in transactions, while sell orders whose price is larger than that of buy orders are kept in a register called a limit order book (LOB) until they are matched or cancelled by their owner. The order flow is treated asynchronously, each new event giving rise either to a new order in the LOB, a new transaction, or a cancellation.

A subspecies of high-frequency traders are market makers who propose buy and sell orders simultaneously. They make money by selling at higher prices than buying; since they have no guarantee to buy and sell the same quantity at about the same times, they also must learn how to adjust their quotes so as to control their inventory and avoid situations where prices move in an adverse way given their inventory. Because traders, and thus market makers, are competing against each other and because the environment is very sensitive to each event, successfully training and deploying a RL agent is hard. 

Let us explain the consequences of the specific difficulties that we face here.

\begin{enumerate}
    
\item {\bf Competitive asynchronous multi-agent systems.}
Learning in competitive systems may lead to unstable dynamics when learning is too fast with respect to the quantity of available information, as seen in numerous agent-based models which undergo a signal-to-noise driven transition between stable and unstable dynamics (see e.g. \cite{chiarella1992,marsili2000exact}). Unstable behaviour includes herding whereby a finite fraction of the population ends up taking the same action repeatedly, which leads to large fluctuations. Asynchronicity encourages imitation, hence herding \citep{mavrodiev2013quantifying}.

\item {\bf Latency and effective events.} 
Latency causes additional headaches for RL: indeed, as there is a delay between the communication of the environment state and the implementation of the action of an agent, the connection between state, action, and reward becomes much looser. 
Even if the data have a nano-second time resolution, in practice market makers suffer significant latencies when acquiring the data, computing their actions, and transmitting them to the market. Latency matters to the type of agents we wish to train: market makers try to minimize their latency by renting computer racks as close as possible to the matching engines of the LOBs and buying access to the fastest networks between two exchanges.

In addition, the choices open to market makers are essentially where to place their limit orders with respect to reference prices. If the latter change, even if the agent keeps the same choice (i.e., action in his point of view), he has to adjust the price of his orders, i.e., possibly cancel or submit new orders. This is what we call effective events (from the point of  view of the system).

\item {\bf Highly sensitive systems with long memory.}
Some systems, and especially multi-agent competitive systems with adaptive agents, may be very sensitive to a single event, for example because there is not much information left and each event may reveal new information, and remember it for a long time.

In financial markets, each order has a long-term impact on future events \citep{lillo2004long,eisler2012price}, which leads to very complex but highly non-random dynamics (see \cite{abergel2016limit,bouchaud2018trades,lehalle2018market} for recent monographies).

\item {\bf Anonymous data for multi-agent systems.}
When there is a single agent interacting with its environment, RL is usually formalized by defining the state of the system (environment and agent)  $s_t$ at time $t$, the (re-)action of the agent, $a_t$, and its reward $R_t$. If the data (or the model) is trustworthy enough, then learning the associations between states, actions, and future rewards is exactly what RL is meant for.

Consider now the case of multi-agent systems. If the data is agent-resolved, i.e., if one knows which agent took which action, the state of each agent can be tracked easily and thus the payoff corresponding to each action can be inferred at least partially. In short, off-line learning is not much different from the single-agent case.

When the multi-agent data are anonymous, the consequences are far reaching: not only one does not know how many agents there are, but one can infer neither their state nor their reward, with precludes the use of inverse RL (see \cite{arora2021survey} for a review). 

In order-driven financial markets, the best publicly available data only reports the orders identification numbers, not their owners, and are therefore anonymous. Academics only rarely have access to trader-resolved data encompassing all the transactions in a given exchange (see however \cite{barber2007aggregate,lillo2015news,sato2023inferring}).

\item {\bf Hard-to-simulate systems. }
The usual way to alleviate data anonymity is to use simulators. Unfortunately, and unsurprisingly, asynchronous, competitive, very reactive multi-agent systems with long memory and latency are hard to simulate. In that case, there may be no model good enough that guarantees that a RL agent will not simply learn to exploit an idiosyncratic weakness of the model and thus have a poor performance when deployed live. 

\end{enumerate}

Still, leveraging RL is very natural in a financial context, as market participants have a clearly defined objective function and the average reaction of the LOB to a single event has a systematic part. For a generic review on the use RL in finance, see \cite{hambly2023RL}. We note however that previous attempts to use RL for a market making agent either used data while neglecting the event-by-event influence of the action of the RL agent on the future stream of events (see e.g. \cite{spooner2018market}), or used LOB simulators, however sophisticated, that only imperfectly mimic real markets (e.g. \citep{oxfordGPU,kumar2020deep}).

Here we propose a generic way to use data in a consistent way for hyper-sensitive,  hard-to-simulate multi-agent systems with anonymous data, taking as an example order driven markets where an agent learns to become a successful market maker. In essence, we advocate consistent data time travel, whereby the time index after the RL agent takes an action chosen so that the effect of the action of the RL agent onto the system is consistent with historical data; this may require to jump to another time with consistent the state before the agent's activity and the resulting effective events of the agents. %This is far from trivial in limit order books where the interaction between the agent and the system has some latency.

\section{Interacting with anonymous multi-agent asynchronous data}

Imagine a system (e.g. a limit order book, excluding the traders) whose state is described by a set of variables $X_{t}$, where the discrete variable $t$ indexes the events. For the sake of simplicity, assume that $X_{t}$ can only
be modified by a event $\alpha_{t}\in\mathcal{A}$, the set of possible events for that particular system (e.g., to place or to cancel an order).

It is important to differentiate between the events (at the system level) and the actions of the agents, because the latter may cause several events and the mapping between the two is context-dependent, as will be made clear in the following.

We write the dynamics of the system as an update equation for $X_t$:
\[
X_{t+1}=F(X_{t,}\alpha_{t}).
\]
In multi-agent systems with agent-resolved data, the event at time $t$ can be attributed to agent $i_t\in\{1,\cdots,N\}$.

Imagine that one wishes to train a RL agent, who receives state $s_t$  and decides to take an action $a_t^\mathrm{RL}$ which results in an effective event $\alpha_t^\mathrm{RL}$. Until now,  the literature  only proposes  two  possibilities: 
\begin{enumerate}
    \item one inserts the RL agent event in the historical  sequence, i.e, one adds $\alpha_t^\mathrm{RL}$ just before the historical action $\alpha_t$
    \item one replaces the historical event $\alpha_{t}$ by the agent's action $\alpha_t^\mathrm{RL}$.
\end{enumerate}

The first possibility is the simplest: it assumes that the insertion of another event does not change the future sequence of orders, hence that the impact of the new event is effectively zero. It is a good scheme for systems which are weakly sensitive to a single  event. 
While it is the usual approach taken by RL papers on market making (e.g. \cite{spooner2018market}), which have to assume that the order size is negligible, limit order books are known to be very sensitive to every single change and do remember them for a long time (see e.g. \cite{eisler2012price}). The biases induced by this assumption are certainly important but hard to quantify.

The second possibility also assumes that the system is only weakly sensitive to the change of a single  event. It however offers a perfect simulation of the reaction of the system if the effective event of the agent is the same one as the historical one, i.e., if $\alpha_t^\mathrm{RL}=\alpha_t$. 
Using historical data naively is therefore only recommended for weakly reactive systems with a short memory. When no good model exist, it is still tempting to use historical data, accounting in some way for the impact of the additional RL agent, but up to now, there is no good way to account remotely realistically for event-by-event impact of the RL agent with historical data.

Here, we propose a third possibility: data time travel. Its aim is to maximise the consistency of the influence of the new RL agent's actions with what happened next in the system, i.e., with the next effective events conditionally on the state (static and dynamic) of the system. To achieve this aim, instead of using historical sequentially, the time index may jump if needed to another point in time that shares both the same system state and the same effective events resulting from actions of the RL agent: consistency is defined at the level of the system with which the agent interacts. In the context of learning to trade or to be market makers, this may sound rather worrying as it breaks causality in a strict sense. However, provided that the state of the market sufficiently encodes the relevant part of history,  it is a better way to use historical data than to completely neglect any influence of the RL agent on the dynamics of the system. In short, to gain local consistency, one must break global causality.

Let us go back to the state time evolution: assume that the state of the system (LOB) is $X_t$ and that the 
historical effective event was $\alpha_{t}$. The RL agent chooses to  play $a_t^\mathrm{RL}$ which results in (effective) action $\alpha_t^\mathrm{RL}$. The next system state is thus given by $X_{t+1}=F(X_t,\alpha_t)$ in historical data and $\tilde X_{t+1}=F(X_t,\alpha_t^\mathrm{RL})$ when the RL agent acts.

One must distinguish two cases: 
\begin{enumerate}
\item $\tilde X_{t+1}=X_{t+1}$: the next state of the system given
by historical data corresponds to the state induced by the action
of the RL agent: the use of historical data is clearly consistent;
\item $\tilde X_{t+1}\ne X_{t+1}$: there may be a large discrepancy between $\tilde{X}_{t+1}$
and $X_{t+1}$ and by extension at later times, i.e., between $\tilde{X}_{t+\tau}$
and $X_{t+\tau}$ with $\tau \ge 1$. 
\end{enumerate}

The latter case is the source all the problems when using historical data naively. Data time travel instead proposes to jump in these cases to another time index which is more consistent with both the system state and the influence of the RL agent onto the system. In other words, one wishes to find $t'$ is such that 
\begin{align}
    X_{t'}&=X_t \label{eq:XtXt}\\
    X_{t'+1}&=\tilde{X}_{t+1} \label{eq:Xt1Xt1}.
\end{align}
Equations \eqref{eq:XtXt} and \eqref{eq:Xt1Xt1} define a consistency criterion between the states at time index $t'$ and $t'+1$ and those induced by the RL agent at time $t$. Accordingly, one should jump to time $t'$ and continue the training of the RL agent from $t'+1$. Note that consistency means that one sometimes needs to jump in to the past, if allowed to do so. If causality is of utmost importance in the training of the RL agent, and if the amount of available data is very large, one can impose $t'>t$. On a side note, we do not impose $t,t'>0$.

There are two potential complications: 
\begin{enumerate}
    \item there are more than one time index that are consistent with the influence of the RL agent. In this case, one can choose uniformly from this set, or impose additional constraints (proximity-based, causal, etc.).
    \item there is no consistent time index, which is the norm if the state is a continuous variable (or a vector that includes continuous variables). The solution is instead to define a distance between two states and to find the indices that minimize it. 
\end{enumerate}

Computing naively the distance between all the time steps is too intensive when the number of data points is reasonably large. This is why we define discrete states and compute dictionaries of the times corresponding to each discrete state. 

A fundamental question is how time travel changes the nature of the system dynamics. In particular, many results in RL literature about the convergence of the agent to an optimal behavior depends on the state of the system being Markovian. Two remarks are in order. First, assuming that the system was Markovian in the first place, time jump does not make it not Markovian: because one jumps to another point in historical data, i.e., to events that actually happened naturally, the nature of the dynamics is unchanged. The second remark is that applying RL to financial markets, known to be non-Markovian systems,  does not aim for learning the optimal behaviour (e.g. policy), but for learning enough to have a profitable agent.

\section{The case of limit order books}

In order-driven markets, traders send wishes to buy or sell preset prices. For example, a buy order $i$ has a price $p_i$ and volume $v_i$; when it is sent to the market, it either finds a pre-existing sell order with a compatible price (lower or equal to $p_i$), or it is stored in the so-called limit order book (LOB). 

Let us introduce some useful notations. At time index $t$,  the best buy orders in the
LOB have a price $b_{1,t}$, the best sell orders have a price $c_{1,t}$ (shortened to $b_t$ and $c_t$),
 while the second buy (sell) prices are denoted by $b_{2,t}$ ($c_{2,t}$). In addition, the total volume of all the orders at $b_{k,t}$ ($c_{k,t}$) is $V^B_{k,t}$ ($V^S_{k,t}$).

While a typical trader may place an
order of a single type (buy or sell) sometimes, market makers usually
display both a buy limit order and a sell limit order at all times,
which ensures that the LOB may be useful both for buyers and sellers
at any given time by containing enough available volume on both buy and sell sides. 

The whole question for a market maker is where to place her orders.
The simplest idea is to place them at the best prices.
If $b_{t}$ and $c_{t}$ are constant, the market maker earns the
spread $\sigma_{t}=c_{t}-b_{t}$ each time someone buys one share and someone else sells one share.
The cumulative amount sold $V^S_{MM,t}$ and amount bought $V^B_{MM,t}$ up to time $t$ are not likely to be equal, thus she accumulates
a net inventory $J_{t}=V^B_{MM,t}-V^S_{MM,t}$ of shares bought and sold, which is risky if the prices move in the wrong direction.

If she places her orders symmetrically with respect to the mid price
$m_{t}=(b_{t}+c_{t})/2$, and if the buy and sell transactions arrive in an unbiased way, $J_{t}$ follows an unbiased
random walk. However, by posting limit orders asymmetrically, e.g.
the sell order at $c_{t}=c_{1,t}$ and the buy order at the second best price \textbf{ $b_{2,t}$},
the MM may control the respective execution probabilities of her orders, thereby indirectly (stochastically) controlling the evolution of her inventory. This is called skewing one's quotes. One order may be placed much beyond the second best price in practice in order to minimize the risk of having an inventory increase in the wrong direction.
When price skewing fails to limit the absolute value of $J_{t}$
and thus the inventory reaches a maximum preset value, the MM is assumed to liquidate her whole inventory (see the literature on optimal market making in stochastic price models \citet{gueant2017optimal,bergault2021closed,barzykin2023algorithmic}).

We assume that each order may be placed either at the best price (denoted
by $1$) or the second best price ($2$), which gives two symmetric
 order placement actions, two asymmetric ones, and a liquidation action
($\emptyset$): the market maker here has five actions to choose from: $a_t \in A=\{1,2\}\times\{1,2\}\cup\{\emptyset,\emptyset\}$,
where the first coordinate denotes the buy side and the second one the sell side and $\times$ is the set outer product.

These are the actions of the RL agent, not to be confused with the effective events in the system which describe what actually happens in the LOB:  one can describe an event in the LOB, restricted to the first two ticks and the spread, by  $\alpha\in\mathcal{A}=\{\delta_{k,\xi}\}_{k=0,1,2;\xi\in\{B,S\}}$, where $\delta$  is the size (signed) of the event, $k$ is the index of the price queue, 1 denoting the best price queue on the $\xi$ side, 2 the second best, and 0 the spread. Note that removing liquidity can be caused either by a trade or by a cancellation. The point is that a single event in the LOB corresponds to a single effective event. 

The mapping between the RL agent actions and the effective events is not unique: the RL agent's actions correspond to a strategy of order placement relative to current best prices. As as consequence, if the agent action changes, or if at least one current best price changes, the RL agent may need to update her orders in the LOB, which results in a series of order cancellation and placement, i.e., a set of effective events. 

Another complication originates from the fact a trader cannot be as fast as the order matching engine, hence, cannot react instantaneously to each tick $t\to t+1$. We need therefore to introduce  the update time index of the RL agent, $u$, with a mapping to the system event index $t(u)$ for all $u$; in practice, $t(u+1)$ is the smallest value of $t$ such that $t(u+1)-t(u)$ is larger than the latency of the agent.  Imagine that the RL agent actions are the same for update indices $u$ and $u+1$, e.g. $a_u=a_{u+1}=\{1,1\}$. Let us focus on the buy side:
\begin{enumerate}
    \item if the best prices did not change at all between two updates, i.e., $b_{t'}=b_{t(u)}$ and $c_{t'}=c_{t(u)}$  for all $t'\in\{t(u),\cdots,t(u+1)\}$, then the MM has nothing to do;
    \item if $b_{t(u+1)}>b_{t(u)}$, the MM has to cancel her buy order at price $b_{t(u)}$ and add a new one at price $b_{t(u+1)}$ so as to keep her buy order at the best buy price. This results in the two effective events $+_{1,B}$ and $-_{2,B}$ if $b_{t(u)}$ equals the second best price at time $t(u+1)$ (and only $+_{1,B}$ otherwise as we focus on the best two quotes on each side) . 
    \item if the buy order of the MM has been fully executed by transactions between two updates, she needs to send a new buy order at $b_{t(u+1)}$, which corresponds to one effective event $+_{1,B}$.
    %if $b_t<b_{t(u)}$ for at least one time $t\in\{t(u),\cdots,t(u+1)\}$ because of transactions, it is safe to assume that the buy order of the MM has been fully executed. As a consequence, the MM has to 
\end{enumerate}
The set of effective events thus must be determined at each $t(u)$ and may be different for all possible actions and LOB changes.

There are several ways to account for latency while training the RL agent. The simplest one is to assume that the RL agent can update its action at most every $l$ seconds ($l$ can be much smaller than 1). This is realistic if the exchange disseminates information with a latency much larger than the total agent latency (communication from the exchange, processing, communication to the exchange), or in the limit of much larger processing time than both communication latencies. A more realistic setting is to consider the inbound latency $l_\textrm{in}$, the processing time $l_\textrm{proc}$, and the outbound latency $l_\textrm{out}$: the RL agent is active every $l_\textrm{proc}$ seconds and receive the LOB state with a delay of $l_\textrm{in}$, and its effective events would therefore be valid with an additional delay of $l_\textrm{out}$. In the following, we use the first setting.

\subsection{Market and agent states}

When active at time $t(u)$, the RL agent receives state $s_u$ which describes both her internal state $\mathcal{I}_u$ and the state of the LOB $X_{t(u)}$, thus $s_u=(\mathcal{I}_u,X_{t(u)})$ 

The state of the LOB $X_t$ should describe the quantities that are relevant to  predict the next LOB state $X_{t+1}$. This a tall order, as the influence of every single event has a long lasting effect. Instead, for the sake of simplicity here, we will approximate $X_t$ with a much simpler description that encodes the most salient features of LOB. What most matter to a market maker is the side (buy/sell) where the next execution takes place, as this information is closely related to how she should skew her quotes in order to adjust their execution probabilities. It is known to depend much on average on the imbalance of the total order size on the best prices
\[
I_{t}=\frac{V_{1,t}^{(B)}-V_{1,t}^{(S)}}{V_{1,t}^{(B)}+V_{1,t}^{(S)}}.
\]
This strongly suggests that the sign of $I_{t}$ (including 0) is the most important bit of information regarding the next trade side. Another instantaneous import quantity in LOB is the spread $\sigma_t=c_t-b_t$, which we will account for.
There are other quantities based on the order flow, i.e., dynamics-related quantities such as the limit order imbalance \citep{cont2014price}) that are important to predict future events. We do not include them in this study.

The state of the agent, $\mathcal{I}_t$, is simplified to the sign of its inventory, including 0. The idea is that the agent should learn how to skew her quotes according to her inventory. 

\subsection{Consistency conditions for data time travel}

The two conditions defined in Eqs \eqref{eq:XtXt} and \eqref{eq:Xt1Xt1} assume that one can describe the system state $X_t$ exactly. Here we will approximate $X_t$ by the quoted volume imbalance sign $I_t$ and the spread actual value $\sigma_t$, thus Eq.\ \eqref{eq:XtXt} becomes
$$X_t\simeq\mu_t=(I_t,\sigma_t)=(I_{t'},\sigma_{t'})=\mu_{t'}\simeq X_{t'}.$$ 

When taken together, both consistency conditions on the states imply that the sets of effective  events caused by the RL agent and of events in historical data  are the same or at the very least compatible. This provides a criterion according to which time jump is necessary. 

Let us denote the set of events caused by the RL agent action $a_t$ by $\mathcal{A}_t^\mathrm{RL}$. The consistency condition in historical data is that the minute dynamics of the LOB after $a_t^{\mathrm{RL}}$ corresponds what would have happened and is the main reason why data time jump improves on the current naive methods. Finding effective events with exactly the same sizes $\delta$ is unlikely, thus we only consider the signs of the effective events. In some cases, historical data is fully compatible with the actions taken by the RL agent. Most of the time, though, one has to jump. 

Note that the effective events may occur in any order in the historical data; in addition, it may be interspersed with the effective events of other traders because of the asynchronous nature of LOBs. As a consequence, we ask for the set of effective events to be found, in any order, within $T_{next}$ time indices of $t'$; if is the case, we take the smallest number of ticks for which historical data after $t'$ includes all the effective events of the RL agent.

Finally, it may happen that there is no $t'$ within the same day that has the same $\mu_t$ and $\{\alpha_{t,a_t}\}$. This most likely happens because we wish to jump to a time with exactly the same spread value. In these cases, we allow some leeway when needed: $t'$ is admissible if 
\begin{equation}
|\sigma_t-\sigma_{t'}|<\Delta \sigma_\textrm{max}\label{eq:Delta_sigma}.
\end{equation}
In practice, we check for compatible jump times with the smallest spread difference in absolute value that still satisfy condition \eqref{eq:Delta_sigma}. We take $\Delta \sigma_\textrm{max}=$0.3\texteuro.

One usually finds more than one candidate jump time. Several restrictions on the jump times can be implemented. For example, one may only allow to jump in the future, in which case the historical data are rapidly exhausted. One also can only allow to jump in a restricted time interval around current time. The risk is to be stuck for a while in a periodic state. We chose to jump to a time chosen at random in the set of candidate times after removing times too close to the current one (10 ticks here). If we do not find any suitable jump time, we simply carry on and jump to $t+|\{\alpha_{t,a_t}\}|$, i.e. the time index incremented by the number of effective events.

% For example, if the MM changes her strategy at time $t^{*}$ from $(1,1)$ to $(1,2)$, the volume on the best sell price will decrease while volume on the second best sell price will increase, while keeping the buy volume unchanged on both the best and the second best buy price; this can be encoded as $a_{t^{*}}=(0,0,-,+)$.

\subsection{Computational speed}

We pre-compute separate dictionaries of indices for a given $I$, or $A$. The set of candidate jump times for a given tuple ($I,A)$ is then simply the intersection between candidates from each dictionary, to which one applies the condition on the spread \eqref{eq:Delta_sigma}. This makes the the distance computation loose its $O(T^2)$ complexity if $T$ is the number of data points ($T\sim 10^5$ here).

The occurrence of a given type of event in a given time interval $[t,t+T_\textrm{next}]$ can encoded by a binary vector (1 if at least one occurs, 0 otherwise). We therefore pre-compute these vectors for all $t'$ within $T_\textrm{next}=20$ events from $t$.

\section{Methods}

\subsection{Data}

This study uses data from BEDOFIH, a high quality data base for academics from Eurofidai. It contains all the needed information to recreate the tick-by-tick history of all the orders sent to several European exchanges. Here we focus on Paris Stock Exchange, more specifically on the most traded asset, Total Energies. We select four arbitrary days  $D=$(2016-01-05 to 2016-01-08). For each day, we keep the full information (order by order) on the two best prices on each side of the LOB, which yields about 200,000 events per day on average.

\subsection{RL setup}

The RL market makers use Q-learning, a model-free approach that is sufficient to explore the influence of the history types. At activation time $t(u)$, the agent receives state $s_u$ and decides where to place her limit orders ($a_u^\mathrm{RL}$). For the sake of simplicity, we restrict the choices to the first two best current limit prices on either side, which yields four action, and the choice to liquidate the inventory. 

The key here is that the payoff is determined by all the events that occur at each tick between $t(u)$ and $t(u+1)-1$. We therefore must accumulate the tick-by-tick rewards between two successive RL agent activation times. The rewards follows the symmetric form of \cite{spooner2018market}; more specifically, if at time $t$, the MM holds an inventory $J_t$, she receives  a payoff
\begin{equation}
    R^{J}_t=J_t\times (m_t-m_{t-1}),
\end{equation}
where $m(t)$ is the mid-price. 
The inventory changes when the orders of the agents are executed, between two activations times, hence total inventory payoff between two activation times is therefore
$$
\mathcal{R}^{J}_{u+1}=\sum_{t'=t(u)+}^{t(u+1)}R^{J}_t.
$$

Between two activation times $t(u)$ and $t(u+1)$, the MM cannot update the price of her orders, which may be therefore  partly or fully filled, by a single transaction or by many of them. To simplify the discussion, let us focus on the buy order of the MM placed at price $b_u^{\mathrm{MM}}$. There are three events that trigger partial or full execution:

\begin{enumerate}
    \item a transaction of volume $V_t^{\mathrm{trans}}$ took place in historical data at time $t$ and price $b_u^{\mathrm{MM}}$; denoting the remaining volume of the buy MM order just before the transaction by $v^B_{\mathrm{MM},t}$ and the volume at that price in historical data by $v^{B}_t$, we assume that the MM order matched volume is proportional to the total volume at the best buy price:  $v^B_{\mathrm{MM},t+1}-v^B_{\mathrm{MM},t}=v^B_{MM,t} v_t/(v^{B}_t+v_{\mathrm{MM},t})$. In other words, we do not 
    \item a transaction took place at time $t$ at a price smaller than $b_{\mathrm{MM},u}$; full execution;
    \item the best sell price is larger or equal to $b_{\mathrm{MM},u}$; full execution.
\end{enumerate}
Each execution of volume $v_{MM}$ entails a reward
\begin{equation}
    R^{B,\mathrm{exec}}_t=v_\mathrm{MM}(m_{t-1}-b_{\mathrm{MM},u}).
\end{equation}
Similarly to the inventory payoff, the total transaction payoff between two activation times is
$$
\mathcal{R}^{B,\mathrm{exec}}_{u+1}=\sum_{t'=t(u)+1}^{t(u+1)}R^{B,\mathrm{exec}}_t.
$$

When the time index reaches $t(u+1)$, the total reward is $\mathcal{R}_{u+1}=\mathcal{R}^{J}_{u+1}\mathcal{R}^{B,\mathrm{exec}}_{u+1}$. The RL agent is trained with $\epsilon$-greedy Q-learning, where the matrix element $Q_{s_u,a_u}$ is updated according to \cite{sutton}
\begin{equation}
    Q_u(s_{u},a_{u})=Q(s_{u-1},a_{u-1})+\beta\left[R_u+\gamma\max_{a'}Q_u(s_{u},a')-Q_u(s_u,a_u)\right]
\end{equation}
and then the action $a_{u+1}$ is taken according to $\arg\max_{a'}Q(s_{u},a')$; with  probability $\epsilon_t$, $a_t$ is instead replaced by a random action (exploration); initially, $\epsilon$ is set to $0.2$ and multiplied by a factor $0.9999$ at each update. Given that there are about 20'000 updates per day, the exploration parameter $\epsilon\simeq 0.05$ at the end of the training period. We fix $\beta=0.001$ and $\gamma=0.97$ (as in \citet{spooner2018market}). The MM posts orders of size 100. If some of her orders are executed between two activation times, she adds more volume to her order at the next activation time. Her maximal inventory is set to 1000 shares: whenever $|J_t(u)|\ge 1000$, she liquidates all her inventory by sending a market order, which entails a half-spread penalty times the inventory size $|J_t(u)|$. Note that we only train a single agent at a time.

\section{Results}

The central question is what differences data time travel makes when training a RL agent with tick-by-tick market data. We expect that a RL agent learns something different in either dynamics  (sequential data or data time travel, shortened to 'jump' in labels and captions thereafter) and that it will show in the average payoff per update. It is not easy to have an intuition regarding which time dynamics leads to the larger payoff in real life nor in the testing periods. However, we do expect that agents train with a given time dynamics will perform less well when tested with the other time dynamics.

For each type of time dynamics and each day, we train $N=48$ RL agents with different random initial $Q$ matrices, which amounts to $392$ agents to train. We use $T_\mathrm{train}=15,000$ agent updates $u$;  we tried several $T_\mathrm{train}$ in the $10000-22500$ range in order to check for overfitting, but did not find any specific pattern of the test average gain as a function of $T_{train}$. This may come from the fact that the agents still explore and learn when tested.

Each agent is tested starting from the $Q$ matrix it had at the end of the training day $d$ for each day $d'$ in the respective testing set (either $d=d'$ or $d\ne d'$).  The performance of agent $i$ trained on day $d$ with time dynamics $n$ and tested on day $d'$ with time dynamics $n'$ is defined as the average payoff per update at time $t$, from the start of the day, denoted by $G^{(d,x)\to(d',x')}_{i,t}$. We first check the gain of the agents on the same day as the training one, ($d=d'$), i.e., measure,
\begin{equation}
    G^{(n)\to(n')}_t=\frac{1}{N}\frac{1}{|D|}\sum_{d\in D}\sum_{i=1}^N G^{(d,n)\to(d,n')}_{i,t},
\end{equation}
then we compute gain cross-validated on the days not used for training, i.e.,
\begin{equation}
    G^{(n)\to(n')}_{\mathrm{cross},t}=\frac{1}{N}\frac{1}{|D|(|D|-1)}\sum_{d\in D}\sum_{d'\ne d \in D}\sum_{i=1}^N G^{(d,n)\to(d',n')}_{i,t}.
\end{equation}
We test each agent in two ways: using historical data  sequentially ($n'=\textrm{seq}$) or with time jumps ($n'=\textrm{jump}$). In this way, we are able to understand the influence of the time dynamics both in the training and testing phases. Although the cross-validation setup is not strictly causal (as $d'<d$ is allowed), our aim is limited to demonstrate that time travel leads to qualitatively different results.

\begin{figure}
\centerline{
\includegraphics[width=0.50\textwidth]{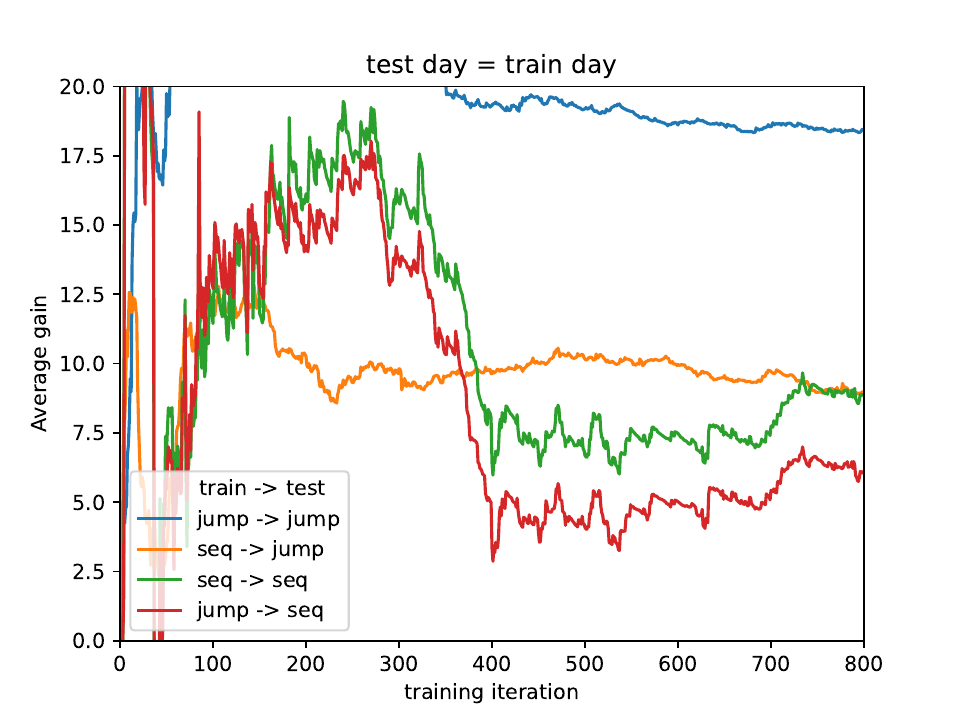}
\includegraphics[width=0.50\textwidth]{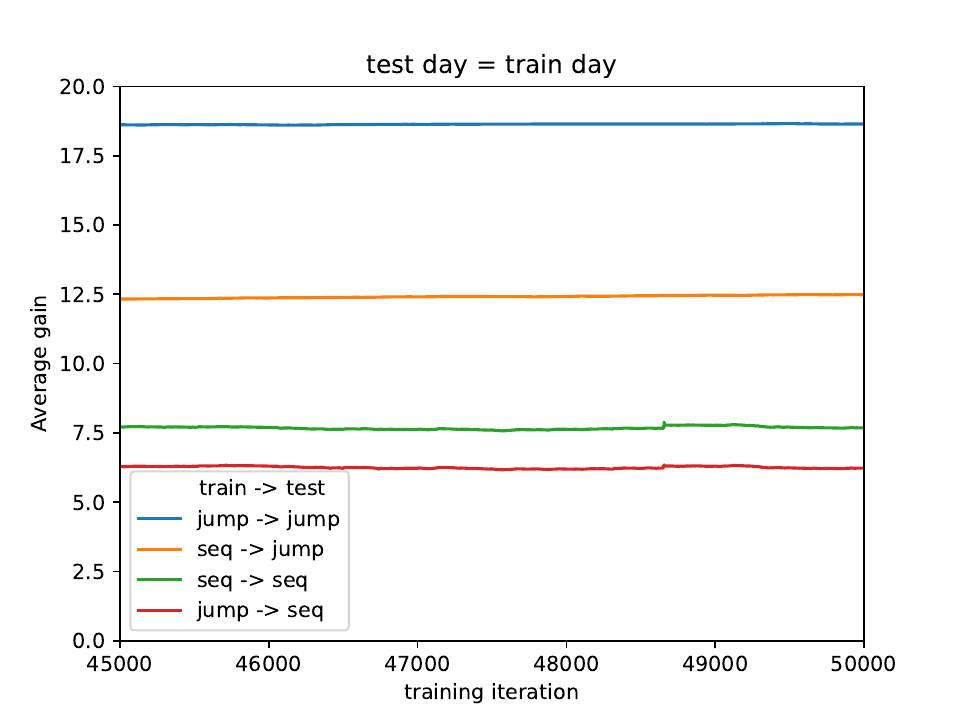}
}

\caption{Average gain of the market maker agents in the test days as a function of number of learning update for various combinations of train / test time evolution scheme; test on the same day as the training, after training. Left plot: zoom on the start of the testing periods; right plot: zoom over the last part of the testing periods. Averages are computed over the 48 agents trained for each day and for each time evolution scheme, over all test days (192); 15,000 training iterations. Data: Total Energies 2016-01-05 to 2016-01-08.\label{fig:gain_-1_0_train}; error bars in the right-hand side plots are within the width of the lines.} 
\end{figure}

The two plots in Fig. \ref{fig:gain_-1_0_train}, summarized in Table \ref{tab:gains_ci}, illustrate our main result: they display $G^{(n)\to(n')}_{t(u)}$ as a function of the number of agent updates; the right hand side plot shows that  agents trained and tested with the same type of time dynamics have a larger gain than those tested on another time dynamics. This means that the type of data does matter and that agents learn something else when fed with different types of time dynamics.

In addition, the agents trained and tested on data time travel (jump) have a larger average gain than agents trained and tested on sequential data. One interpretation is that following the a more consistent historical dynamics yields larger gains when training and testing RL agents. In other words, assuming neglecting ones impact on the actual dynamics wrongly suggests that using RL for market making is harder than it actually is. 

\begin{figure}
\centerline{
\includegraphics[width=0.50\textwidth]{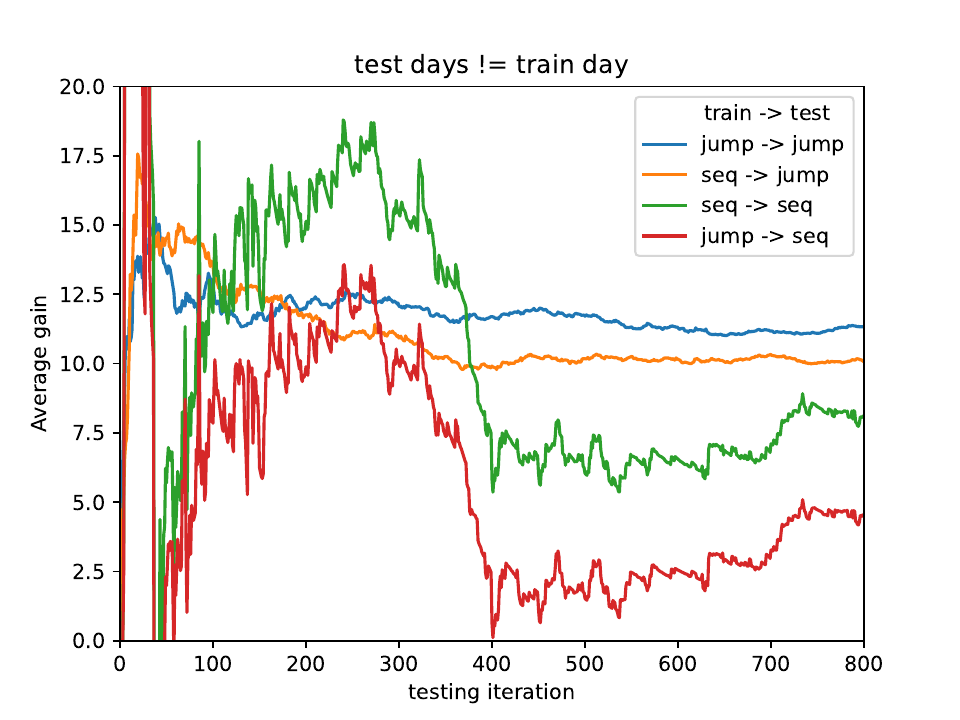}
\includegraphics[width=0.50\textwidth]{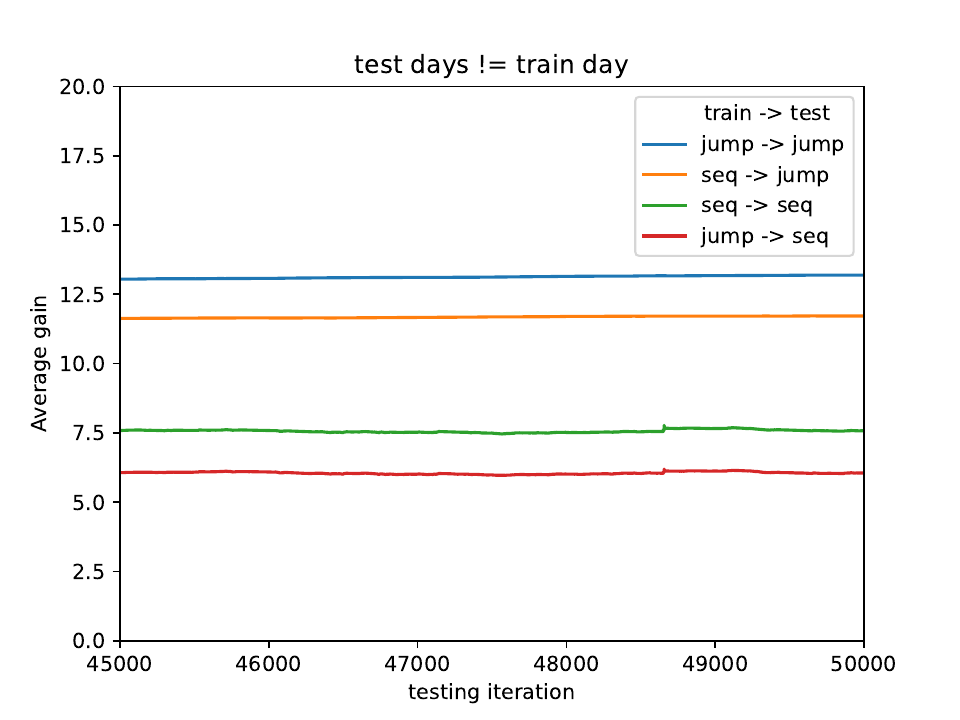}
}
\caption{Average gain of the market maker agents in the test days as a function of number of learning update for various combinations of train / test time evolution scheme; test on the same day as the training, after training. Left plot: zoom on the start of the testing periods; right plot: zoom over the last part of the testing periods. Averages are computed over the 48 agents trained for each day and for each time evolution scheme, over all test days (576); 15,000 training iterations. Data: Total Energies 2016-01-05 to 2016-01-08.\label{fig:gain_-1_0_test}} 
\end{figure}

The cross-validated average gain per update (Fig. \ref{fig:gain_-1_0_test} and Table \ref{tab:gains_ci})tells exactly the same story, albeit with different average gains. One notes that the gains of the agents tested on sequential data does not change much if tested on the say day as the training phase ($d=d'$) or otherwise ($d\ne d'$), which means that while they achieve a positive gain (at least in the data used and given all the backtest hypotheses, including no queue priority), they do not learn much specific to each day. On the contrary, testing the agents with time jumps yields significantly smaller gains than with the train data, especially for the agents trained with time jumps. This is consistent with the idea that consistent historical trajectories, i.e., using data time travel, contains more information.

\begin{table}
\begin{tabular}{cllll}\toprule
 & jump $\to$ jump & seq $\to$ jump & seq $\to$ seq & jump $\to$ seq\\
\midrule
test on train ($d=d'$)& 18.6(0.2) & 12.5(0.2) & 7.7(0.2) & 6.2(0.2)  \\ 
cross-validated test ($d\ne d'$)&  13.2(0.1) & 11.7(0.1) & 7.6(0.1) & 6.1(0.1)  \\
\bottomrule
\end{tabular}
\caption{Average gain between agent update for various train and test time dynamics combinations. Test on train refers to gains measured on the same day as the training day, while cross-validated gain is computed on all the days different from the training day, averaged over all the training day. The parenthesis refer to standard deviations of the  gain averaged over all the test days and for all the 48 agents for each calibration date $d$ (192 samples for $d=d'$ and 576 for $d\ne d'$). Same parameters as in Fig. \ref{fig:gain_-1_0_train}.\label{tab:gains_ci}}
\end{table}

\begin{figure}
\centerline{
\includegraphics[width=0.50\textwidth]{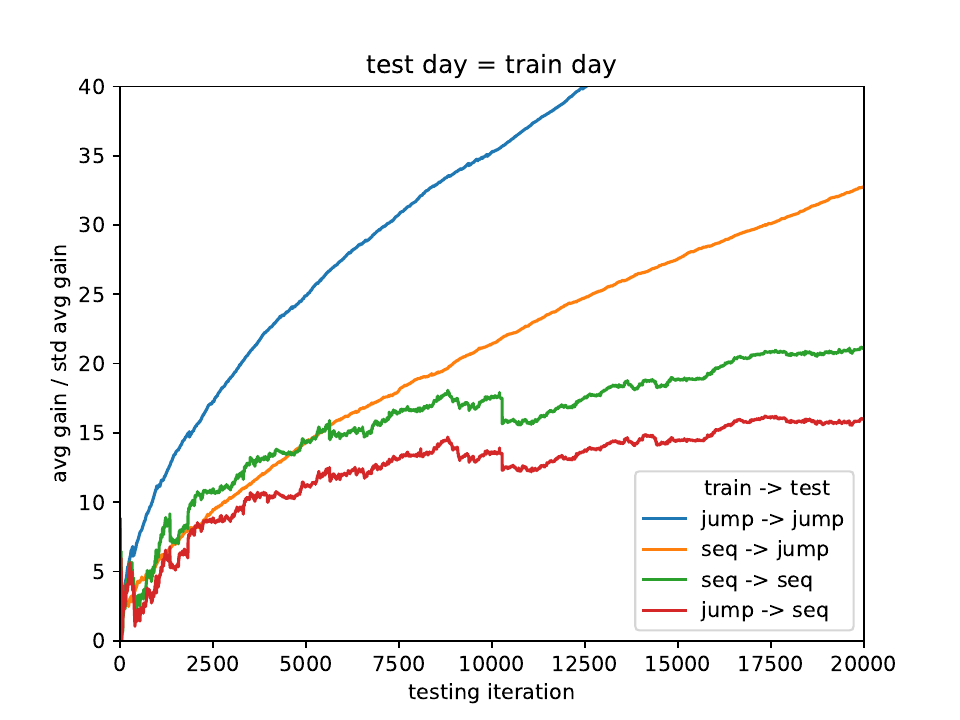}
\includegraphics[width=0.50\textwidth]{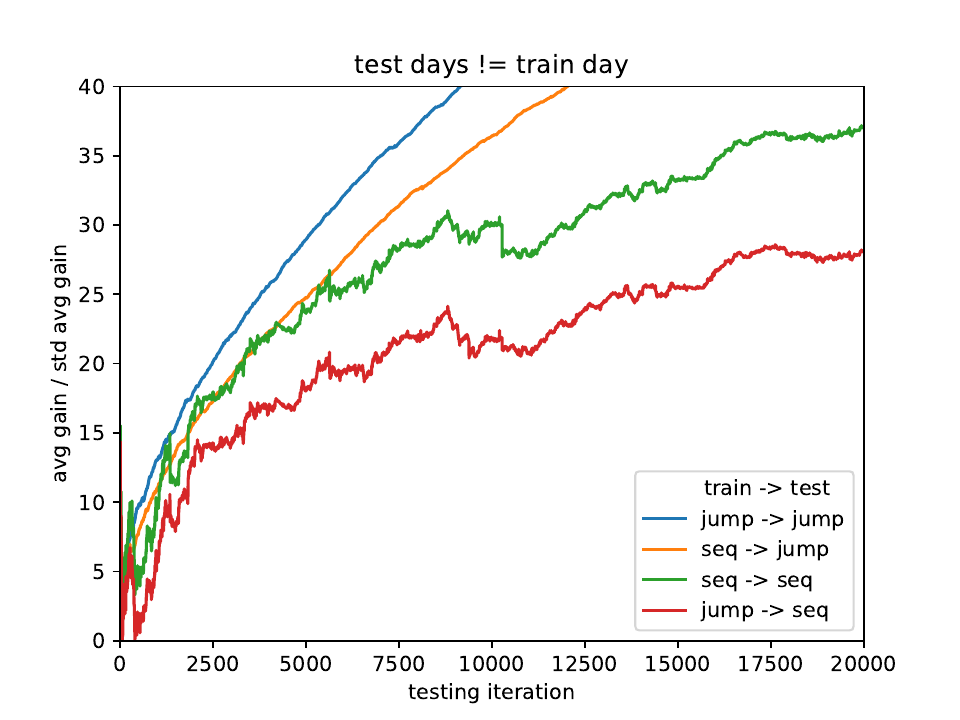}
}

\caption{Average signal-to-noise ratio of the gain of market maker agents as a function of the number of learning updates in the test phase for various combinations of train / test time evolution scheme; test on the same day as the training, after training. Left plot: testing on the same day as training; right plot: testing on non-training days.  Averages are computed over the 48 agents trained for each day and for each time evolution scheme, over all test days (192 when test day $=$ train day, 576 otherwise); 15,000 training iterations. Data: Total Energies 2016-01-05 to 2016-01-08.\label{fig:SNR-1_0_train_test}} 
\end{figure}

Finally, we compute the signal-to-noise ratio of the reward per update (Fig. \ref{fig:SNR-1_0_train_test}), which is mechanically larger when there are more test days (right hand side plot). One first sees that using data time travel yields a better risk-adjusted reward when train in the same way, and eventually for the agents trained with sequential data. For intermediate times, the signal-to-noise ratio is essentially the same for this kind of agents when tested on either sequential data or with data time travel. However, during the first $\sim$5000 updates, the signal-to-noise ratios of their gains are the same for either time dynamics in the test phase; we interpret this as the time needed to learn how to exploit the new type of time dynamics 

\section{Discussion}

Using anonymous data for offline RL in a competitive multi-agent system is possible, provided that one cares to use them in a consistent way, which naturally suggests to exploit data time travel in order to find consistent trajectories. While we used limit order books as a textbook example of such systems, the validity of our approach is not restricted to financial markets. Market making in limit order books provided indeed an ideal application of data time travel. Compared to the usual naive zero-impact  sequential way, data time travel yields a better signal-to-noise ratio and a more consistent loss of gains between the test and the train periods. The importance of taking one's impact, however small, in a competitive multi-agent competitive system was investigated in simple games, which clearly shows that it leads to qualitatively different global and individual behaviour, incidentally by also increasing the signal-to-noise ratio of agent payoffs \citep{marsili2000exact,challet2004minority}.

The clear difference of outcomes between agents trained and tested on either sequential data or with date time travel shows that the agents learn a different dynamics. Using  data in a more consistent way implies that the RL agents learn more realistic paths, i.e., paths closer to those that the system would have taken if the agents had been playing against it. This is probably why the gain is is larger for agents trained and tested on consistently jumping data. The fact that the payoff is larger when using data time travel also suggests that  wrong conclusions about the difficulty of training market maker with RL may be inferred, for example the difficulty of achieving positive payoffs. 

We have also emphasized the importance of accounting for the facts that the actions of a RL agent do not translate into the same set of effective events on the system, and that accounting explicitly for the influence of the different types of latency make the whole problem of consistent dynamics more intricate.  

 While conceptually data time travel makes better use of the available data, its level of consistency  depends on how well the system dynamics is encoded by the chosen system state encoding and influence matching. For example, we neglected the dynamics of the system before the current time when looking for compatible jump times, which can be remedied in a straightforward way for example by include an encoding of the last few events in the limit order book.

More generally, we chose a fairly simplified setup here which has the advantage of speed and was sufficient to demonstrate the validity of our approach. A more realistic setup but much slower would require to use continuous variables and computing distances between dim indices, with partial and possibly expanding index sampling.

\bibliographystyle{plainnat}
\bibliography{biblio}

\end{document}